\documentstyle[12pt,A4]{article}
\begin{document}
\label{zacatek-vys1_99}
\noindent
\hspace*{8cm}{\sc acta univ. palacki. olomuc.,}\\
\hspace*{8cm}{\sc fac.\,rer.\,nat.\,{\small (1999)},\,physica\,{\small
38}},\\
\hspace*{8cm}{\small \pageref{zacatek-vys1_99} -- \pageref{konec-vys1_99}}\\
\hspace*{8cm}\rule[3mm]{6.2cm}{0.2mm}\\[1cm]
\centerline{\large \bf FINAL GENERALIZATION OF THE THREE}\\
\centerline{\large \bf COUPLED OSCILLATORS MODEL IN THE CRYSTAL}\\
\centerline{\large \bf OPTICAL ACTIVITY}\\[2mm]
\centerline{\bf Ivo Vy\v s\'\i n, Kamila Sv\' a\v ckov\' a} \\[3mm]
\noindent {\small Department of Theoretical Physics, \ Natural Science
Faculty of Palack\' y University, \ Svobody 26, {\bf 771~46 Olomouc}, \
Czech Republic}\\[4mm]
\centerline{\it Received 12th February 1999}
\vspace*{0.5cm}

{\sc ABSTRACT:} \ In this paper we generalize the quantum mechanical
model of three coupled oscillators because of its usage in the crystal optical
activity. Using this model we can include the influence of all
essential couplings between single oscillators which represent the
molecules or atoms of optically active crystals belonging for
example to the space groups of symmetry $D_{3}^{4}$ or $D_{3}^{6}$.
The single oscillators are damped and therefore we can include
both parts of the optical activity - optical rotatory
dispersion and circular dichroism - into computations. We present more universal formulas
for description of the above mentioned parts.
\section{Introduction}
\label{uvod}
\hspace \parindent
In this paper we want to generalize the results of the previous paper
\cite{Vsv} in which we solved the optical rotatory
dispersion (ORD) of crystals belonging to
the space groups of symmetry $D_{3}^{4}$ or $D_{3}^{6}$. The theory
was based on the specific quantum mechanical model of three coupled
linear harmonic oscillators representing
the elementary cell of these crystals as the
smallest structure system. But the model used in \cite{Vsv} was solved
for the undamped single oscillators forming the compound oscillator and
therefore this model described ORD only as one part of the optical
activity (OA) and the results were valid only for the wavelength range
outside the absorption band.
Now we can use the model of three damped coupled oscillators and in this
way we
also describe the second part of the OA - circular dichroism (CD) - and
remove the limitation of the wavelength range. But
as will be shown, we need the results of the paper \cite{Vsv} for
the solution this generalized model.

ORD is the dependence of the rotation of the linear polarized light per
unit length on the frequency or the wavelength. CD is the ellipticity
of the outgoing wave from the optically active medium. It is based on the
fact that the right and the left circularly polarized waves, in which the
linear polarized wave is decomposed, and which propagate in optically active
medium with different phase velocities, are absorbed differently. OA is
fully described by the complex rotatory power $\bar {\rho }$ that is given
by the relation
\begin{equation}
\label{01}
\bar {\rho }=\rho +i\sigma =\frac{\omega }{2c}\left (
\bar{n}_{l}-\bar{n}_{r}\right)=\frac{\omega}{2c}\left[ \left(
n_{l}-n_{r}\right)+i\left( \kappa _{l}-\kappa _{r}\right) \right],
\end{equation}
where $\rho $ is the ORD and $\sigma $ is the CD, $\bar{n}_{l}$ and
$\bar{n}_{r}$ are the complex refractive indices of medium for the left
and the right circularly polarized waves. $n_{l}$, $n_{r}$ and $\kappa
_{l}$, $\kappa _{r}$ are the real and the imaginary parts of complex
refractive indices.

It is well known that the molecules or atoms
forming the crystals of the space groups of symmetry $D_{3}^{4}$ and
$D_{3}^{6}$ lie on the helix with screw axis of symmetry and
their optical activity is  caused by
interactions among molecules or atoms in the most important cases.
This is true (with the exception of
camphor) for all typical representative crystals belonging to these
space groups. In the paper \cite{Vsv} we gave reasons for
using the three
coupled oscilators model. Here we mention again that this model fully
represents the elementary cell of the crystals belonging to the space
groups of symmetry $D_{3}^{4}$ and $D_{3}^{6}$ and it also enables
to include
the couplings between even and odd oscillators on the helix. In
\cite{RiV} we have proved that these couplings also have the substantial
influence on the OA in the case when the diameter of the helix is
comparable with the height of the elementary cell or smaller than its
height. But it is known that it is valid for all important crystals with
the screw axis of symmetry.

Our model of three coupled oscillators is the same as in \cite{Vsv}.
The oscillators lie
on the helix, which axis (the crystal axis $c$) is parallel with
the $z$ coordinate axis. The second oscillator lies in the plane $z=0$,
its coordinates are $x,y,0$ and its direction of vibrations is described
by direction cosines $\cos\alpha ,\cos\beta ,\cos\gamma$.
The first oscillator lies
on the helix in the plane $z=-d$ and its direction of vibrations is
turned by the angle $-\theta $ around the $z$ axis with respect to the second
one. Similarly the third oscillator lies on the helix in the plane $z=d$
and its direction of vibrations is turned by the angle $+\theta $ around
$z$ axis with respect to the second one. The angle $\theta $ is of course
$120 \hspace*{1mm} \mbox{deg}$ for the crystals belonging to the space groups of symmetry
$D_{3}^{4}$ and $D_{3}^{6}$. We suppose that all oscillators are
identical and that we can neglect the influence of other helices
which is given by the character of couplings between adjacent
oscillators on one helix and between oscillators belonging to the
different helices. Further we study the OA only in the direction of the
crystal axis where OA is not screened by birefringence.

Further it should be noted that the problems of crystalline OA (or ORD
or CD) were also solved by using other methods.
Here we can mention for example the exciton theory
\cite{Agr1,Agr2,Tsv,Kato}, Lagrangian formalism \cite{Nels}, Schwinger
scattering \cite{BF} etc. But we think that the application of coupled
oscillators models in the crystal optical activity is still important
because the results are convenient for good approximation of the OA
experimental data. This is also the most important aim of this paper.
We want to get a new extension of the formulas for description ORD and
CD experimental data.

\section{Crystal OA based on the three coupled oscillators model}
\label{quant}
\hspace\parindent
We can set the Hamiltonian for one compound oscillator which contains
three coupled oscillators in the field of the left and the right
circularly polarized waves
\begin{eqnarray}
\label{1}
\hat {\cal H}^{l,r}\Psi &=&-\frac{\hbar ^{2}}{2m}\sum_{\xi =1}^{3}\frac{\partial
^{2}\Psi }{\partial
r_{\xi }^{2}}+\frac{m\omega _{0}^{2}}{2}\sum_{\xi =1}^{3}r_{\xi
}^{2}\Psi \nonumber \\
&&+\left[\mu _{1}(r_{1}r_{2}+r_{2}r_{3})+\mu _{2}r_{1}r_{3}\right ]\Psi
+\frac{i}{m\omega }\sum_{\xi =1}^{3}F_{\xi }^{l,r}\hat {p}_{\xi
}e^{\gamma _{0}t}\Psi ,
\end{eqnarray}
where $r_{1},r_{2},r_{3}$ are the displacements of the oscillators from
their equilibria, $\mu _{1}(r_{1}r_{2}+r_{2}r_{3})+\mu _{2}r_{1}r_{3}$ is the
potential energy of mutual interactions of the oscillators and
$F_{\xi }^{l,r}=eE_{\xi }^{l,r}$; $\xi =1,2,3$ are the electric forces
projections of the left and the right circularly polarized wave into the
vibration directions of the oscillators, $e$ is the charge of electron,
$E_{\xi}^{l,r}$ is the intensity of the electric field.
The upper indices $l,r$ hold for the left and the right circularly polarized
waves, $\hat p_{\xi }$ are the momentum operators of all single
oscillators. The small positive parameter $\gamma _{0}$ gives the
possibility of the adiabatic interaction at the time $t=-\infty$. The
damping following from the limited lifetime of oscillators in their
excited states is formally introduced by the parameter $\gamma _{0}$.

The electric vectors $\vec{E}^{l,r}$ of the left and the right circularly
polarized waves propagating along $z$ axis have the components
\begin{eqnarray}
\label{2}
E_{x}^{l}=E_{0}e^{-i\left(\omega t-k_{l}z\right)},\qquad \qquad
E_{y}^{l}=E_{0}e^{-i\left(\omega t-k_{l}z-\frac{\pi}{2}\right)},\nonumber
\\
E_{x}^{r}=E_{0}e^{-i\left(\omega t-k_{r}z\right)},\qquad \qquad
E_{y}^{r}=E_{0}e^{-i\left(\omega t-k_{r}z+\frac{\pi}{2}\right)}
\end{eqnarray}
and the vector $\vec{E}^{l,r}$ projections of these waves into the directions
of vibrations of all oscillators are
\begin{eqnarray}
\label{3}
E_{1}^{l,r}&=&E_{0}e^{-i\omega t}\left[ \left(\alpha \cos \theta +\beta
\sin \theta \right)e^{-i\phi _{l,r}}+\left(-\alpha \sin \theta +\beta
\cos \theta \right)e^{i\left(-\phi _{l,r}\pm
\frac{\pi}{2}\right)}\right],
\nonumber \\
E_{2}^{l,r}&=&E_{0}e^{-i\omega t}\left( \alpha +\beta e^{\pm
i\frac{\pi}{2}}\right),\\
E_{3}^{l,r}&=&E_{0}e^{-i\omega t}\left[ \left(\alpha \cos \theta -\beta
\sin \theta \right)e^{i\phi _{l,r}}+\left(\alpha \sin \theta +\beta
\cos \theta \right)e^{i\left(\phi _{l,r}\pm
\frac{\pi}{2}\right)}\right]\nonumber
\end{eqnarray}
where the values $\phi _{l,r}=k_{l,r}=\bar{n}_{l,r}\omega d/c$ are
complex phase shifts for the left (index $l$) and for the right (index
$r$) circularly polarized waves. In all terms in eq. (\ref{3}) and
further the $+$ sign holds for the left and the $-$ sign for the right
circularly polarizad waves.

Due to the mutual interaction the natural frequency $\omega _{0}$ splits
into three adjacent frequencies. We can solve our problem of three coupled
oscillators using the normal coordinates $q_{1},q_{2},q_{3}$ which we
can easy derive for the system of three classical coupled oscillators.
If we have derived the relations for normal coordinates we can
express the $r_{1},
r_{2},r_{3}$ as combinations of these normal coordinates that are
\begin{eqnarray}
\label{4}
r_{1}&=&\frac{1}{\sqrt{2+A_{1}^{2}}}q_{1}+\frac{1}{\sqrt{2}}
q_{2}+\frac{1}{\sqrt{2+A_{3}^{2}}}q_{3},\nonumber \\
r_{2}&=&\frac{A_{1}}{\sqrt{2+A_{1}^{2}}}q_{1}+\frac{A_{3}}
{\sqrt{2+A_{3}^{2}}}q_{3},\\
r_{3}&=&\frac{1}{\sqrt{2+A_{1}^{2}}}q_{1}-\frac{1}{\sqrt{2}}q_{2}
+\frac{1}{\sqrt{2+A_{3}^{2}}}q_{3}\nonumber
\end{eqnarray}
where
\begin{equation}
\label{5}
A_{1}=\frac{-\mu _{2} +\sqrt{\mu _{2}^{2}+8\mu _{1}^{2}}}{2\mu _{1}},
\qquad
A_{3}=\frac{-\mu _{2}-\sqrt{\mu _{2}^{2}+8\mu _{1}^{2}}}{2\mu _{1}}.
\end{equation}
After substituting (\ref{4}) into (\ref{1}) we get
\begin{eqnarray}
\label{6}
\hat {\cal H}^{l,r}\Psi &=&-\frac{\hbar ^{2}}{2m}\sum_{\eta =1}^{3}\frac{\partial ^{2}
\Psi }{\partial q_{\eta }^{2}}+\frac{m{\omega }_{0}^{2}}{2}
\sum_{\eta =1}^{3}q_{\eta }^{2}\Psi -\frac{1}{2}\left (\mu
_{1}A_{3}q_{1}^{2}+\mu _{2}q_{2}^{2}+\mu _{1}A_{1}q_{3}^{2}\right )\Psi
\nonumber\\
&&+\frac{i}{m\omega }\biggl[ \frac{1}{\sqrt{2+A_{1}^{2}}}\left
(F_{1}^{l,r}+A_{1}F_{2}^{l,r}
+F_{3}^{l,r}\right )
\hat {p}_{q_{1}} +\frac{1}{\sqrt{2}}\left (F_{1}^{l,r}-F_{3}^{l,r}
\right )\hat {p}_{q_{2}} \nonumber \\
&&+\frac{1}{\sqrt{2+A_{3}^{2}}}\left (F_{1}^{l,r}+A_{3}F_{2}^{l,r}
+F_{3}^{l,r}\right )\hat {p}_{q_{3}}\biggr]e^{\gamma _{0}t}\Psi.
\end{eqnarray}

It may be simply verified that the expressions
$\frac{1}{\sqrt{2+A_{1}^{2}}}(F_{1}^{l,r}+\sqrt{2}F_{2}^{l,r}+F_{3}^{l,r}),
\frac{1}{\sqrt{2}}(F_{1}^{l,r}-F_{3}^{l,r})$ and
$\frac{1}{\sqrt{2+A_{3}^{2}}}(F_{1}^{l,r}-\sqrt{2}F_{2}^{l,r}+F_{3}^{l,r})$
in the last equation are the
electric forces in three normal coordinates. We denote them as
$F_{q_{1}}^{l,r},F_{q_{2}}^{l,r}$ and $F_{q_{3}}^{l,r}$. The equation
(\ref{6}) can be separated now by means
\begin{equation}
\label{7}
\Psi (q_{1},q_{2},q_{3},t)=\Psi _{1}(q_{1},t)\Psi _{2}(q_{2},t)\Psi
_{3}(q_{3},t),
\end{equation}
\begin{equation}
\label{8}
\hat {\cal H}^{l,r}=\hat {\cal H}^{l,r}_{q_{1}}+\hat {\cal
H}^{l,r}_{q_{2}}+\hat {\cal H}^{l,r}_{q_{3}}.
\end{equation}

Substituting (\ref{7}) and (\ref{8}) into (\ref{6}) we get three
equations that are
\begin{eqnarray}
\label{9}
\hat {\cal H}^{l,r}_{q_{\eta }}\Psi_{\eta }(q_{\eta },t)&=&-\frac{\hbar
^{2}}{2m}\frac{\partial ^{2}\Psi _{\eta }(q_{\eta },t)}{\partial
q_{\eta }^{2}}+\frac{m\omega _{\eta }^{2}}{2}q_{\eta }^{2}\Psi _{\eta
}(q_{\eta },t)\nonumber \\
&&+\frac{i}{m\omega }\left(
F_{q_{\eta }}^{l,r}\hat{p}_{q_{\eta }}\right)e^{\gamma _{0}t}\Psi
_{1}(q_{\eta },t),\quad \eta =1,2,3
\end{eqnarray}
where
\begin{eqnarray}
\label{10}
\omega _{1}^{2}&=&\omega _{0}^{2}+\frac{\mu _{2}+\sqrt{\mu _{2}^{2}+8\mu
_{1}^{2}}}{2m}, \nonumber \\
\omega _{2}^{2}&=&\omega _{0}^{2}-\frac{\mu _{2}}{m},\\
\omega _{3}^{2}&=&\omega _{0}^{2}+\frac{\mu _{2}-\sqrt{\mu _{2}^{2}+8\mu
_{1}^{2}}}{2m}. \nonumber
\end{eqnarray}

We see that the natural frequencies of single oscillators $\omega _{0}$
are split into three frequencies of the normal modes of vibrations.

If we use the eq. (\ref{3}) we can derive for the electric forces in the
normal coordinates the expression
\begin{equation}
\label{11}
F_{q_{\eta }}^{l,r}=e\left (a_{q_{\eta }}^{l,r}\right
)E_{0}e^{-i\left(\omega t+\sigma _{q_{\eta }}^{l,r}\right)}
\end{equation}
and the coefficients $a_{q_{\eta }}^{l,r}$ are given by the relations
\begin{eqnarray}
\label{12}
\left (a_{q_{1}}^{l,r}\right )^{2}&=&\frac{\alpha ^{2}+\beta
^{2}}{2+A_{1}^{2}}\left
[A_{1}^{2}+4A_{1}\cos \theta +4\cos ^{2}\theta \pm 4\phi _{l,r}\sin
\theta \left (-A_{1}-2\cos \theta \right )\right ],\nonumber \\
\left (a_{q_{2}}^{l,r}\right )^{2}&=&2(\alpha ^{2}+\beta ^{2})\left
(\sin \theta \pm 2\phi _{l,r}\cos \theta \right )\sin \theta,\\
\left (a_{q_{3}}^{l,r}\right )^{2}&=&\frac{\alpha ^{2}+\beta
^{2}}{2+A_{3}^{2}}\left
[A_{3}^{2}+4A_{3}\cos \theta +4\cos ^{2}\theta \pm 4\phi _{l,r}\sin
\theta \left (-A_{3}-2\cos \theta \right )\right ];\nonumber
\end{eqnarray}
$\sigma _{q_{\eta }}^{l,r}$ only have the meaning of the phase shifts.
We can see that the Hamiltonian has a nonperturbed part
\begin{equation}
\label{13}
\hat {\cal H}_{q_{\eta }}^{0}=-\frac{\hbar ^{2}}{2m}\frac{\partial
^{2}}{\partial q_{\eta }^{2}}+\frac{m\omega _{\eta }^{2}}{2}q_{\eta
}^{2}
\end{equation}
which depends only on the index $\eta $ of normal mode and does not
depend on the polarization of the light wave. On the other hand the
perturbed part of the Hamiltonian
\begin{equation}
\label{14}
\hat {\cal H}_{q_{\eta }}^{{l,r}^{p}}=\frac{ie}{m\omega }\left( a_{q_{\eta
}}^{l,r}\right) E_{0}\hat {p}_{q_{\eta }}e^{-i\left(\omega
t+\sigma_{q_{\eta }}^{l,r}\right)+\gamma _{0}t}
\end{equation}
depends also on the polarization of the wave.

The mean value of the electric dipol moment induced by the
left and the right circularly polarized wave that we hold as a small
perturbation we can solve by means of the Kubo theorem \cite{Kubo}. We
get the relation for the mean value of the dipol moment
\begin{equation}
\label{15}
\left< \overline {e\left( a_{q_{\eta }}^{l,r}\right) q_{\eta
}}\right>=\left< e\left( a_{q_{\eta }}^{l,r}\right) q_{\eta }
\right> _{\eta _{0}}+\frac{ie^{2}}{m\omega }\left( a_{q_{\eta
}}^{l,r}\right) ^{2}E_{0}\left< \left< \hat {q}_{\eta },\hat
{p}_{q_{\eta }}\right> \right> _{\omega }e^{-i\left( \omega t+\sigma
_{q_{\eta }}^{l,r}\right) +\gamma _{0}t},
\end{equation}
but the first term on the right side of this equation is the constant
dipole moment of the system and it has no meaning for us. The expression
$\left< \left< \hat{q}_{\eta },\hat{p}_{q_{\eta }}\right> \right>
_{\omega }$ is the Fourier transform of the retarded Green function
$\left< \left< \hat{q}_{\eta },\hat{p}_{q_{\eta }}\right> \right> _{t}$
of operators $\hat{q}_{\eta }$ and $\hat{p}_{q_{\eta }}$, that is
\begin{equation}
\label{16}
\left< \left< \hat{q}_{\eta },\hat{p}_{q_{\eta }}\right> \right>
_{\omega }=\frac{1}{\hbar }\int \limits_{-\infty }^{\infty }e^{i\omega
t-\gamma _{0}t}\left< \left< \hat{q}_{\eta },\hat{p}_{q_{\eta }}\right>
\right> _{t}dt.
\end{equation}

For the retarded Green function $\left< \left< \hat{q}_{\eta
},\hat{p}_{q_{\eta }}\right> \right> _{t}$ in the ground states of
quantum system $|\eta _{0}\rangle $ we hold the relation
\begin{equation}
\label{17}
\left< \left< \hat{q}_{\eta
},\hat{p}_{q_{\eta }}\right> \right> _{t} =-i\vartheta (t)\left< \eta
_{0}\left| \left[ \hat{\tilde q}_{\eta }(t),\hat{p}_{q_{\eta }}\right]
\right|\eta _{0}\right>.
\end{equation}

In the eq. (\ref{17}) $\vartheta (t)$ is the unit step function,
$\hat{\tilde q}_{\eta }(t)$ is the operator of $q_{\eta }$ in the
interaction representation
\begin{equation}
\label{18}
\hat{\tilde q}_{\eta }(t)=e^{i\hat {\cal H}_{q_{\eta }}^{0}t/\hbar
}\hat{q}_{\eta }e^{-i\hat {\cal H}_{q_{\eta }}^{0}t/\hbar
}
\end{equation}
and for the application of the operator in the exponent on the wave
function holds
\begin{equation}
\label{19}
e^{\pm i\hat {\cal H}_{q_{\eta }}^{0}t/\hbar }|\eta _{n}\rangle =e^{\pm
iE_{\eta _{n}}t/\hbar }|\eta _{n}\rangle =e^{\pm i\omega _{\eta
_{n}}t}|\eta _{n}\rangle .
\end{equation}

In following rearrangement we use that for matrix elements of the
production of any operators $\hat{F}_{q_{\eta }}$ and $\hat{K}_{q_{\eta
}}$ holds in the normal modes
\begin{equation}
\label{20}
\left< \eta _{m}\left| \hat{F}_{q_{\eta }}\hat{K}_{q_{\eta }}\right|
\eta _{n}\right>=\sum_{k}\left< \eta _{m}\left| \hat{F}_{q_{\eta
}}\right| \eta _{k}\right> \left< \eta _{k}\left| \hat{K}_{q_{\eta
}}\right| \eta _{n}\right>
\end{equation}
and further we use that the matrix elements of operator
$\hat{p}_{q_{\eta }}$ are
\begin{equation}
\label{21}
\left< \eta _{m}\left| \hat{p}_{q_{\eta }}\right| \eta _{n}\right>
=im\omega _{\eta _{mn}}\left< \eta _{m}\left| \hat{q}_{\eta }\right|
\eta _{n}\right >;
\end{equation}
$\omega _{\eta _{mn}}=\omega _{\eta _{m}}-\omega _{\eta _{n}}$.

Now we can solve that
\begin{equation}
\label{22}
\left< \eta
_{0}\left| \left[ \hat{\tilde q}_{\eta }(t),\hat{p}_{q_{\eta }}\right]
\right|\eta _{0}\right>=im \sum_{k}\omega _{\eta _{k0}}\left| \left<
\eta _{k}\left| \hat{q}_{\eta }\right|\eta _{0} \right> \right|^{2}\cdot
\left( e^{-i\omega_{\eta _{k0}}t}+e^{i\omega _{\eta _{k0}}t}\right)
\end{equation}
and after substituting the eq. (\ref{22}) into (\ref{17}) and solving
the integral on the right side of eq. (\ref{16}) we have
\begin{equation}
\label{23}
\left< \left< \hat{q}_{\eta },\hat{p}_{q_{\eta }}\right> \right>
_{\omega }=\frac{2i\omega m}{\hbar }\sum_{k}\frac{\omega _{\eta
_{k0}}\left| \left< \eta _{k}\left| \hat{q}_{\eta }\right| \eta
_{0}\right> \right| ^{2}}{\omega ^{2}-\omega _{\eta _{k0}}^{2}+2i\gamma
_{0}\omega }.
\end{equation}

We can solve the relation for the induced dipole moments $d_{q_{\eta }}^{l,r}$ by means
of (\ref{23}) and (\ref{15})
\begin{equation}
\label{24}
d_{q_{\eta }}^{l,r}=\sum_{k}\frac{2e^{2}\left( a_{q_{\eta
}}^{l,r}\right) ^{2}\omega _{\eta _{k0}}\left| \left< \eta _{k}\left|
\hat{q}_{\eta }\right| \eta _{0}\right> \right| ^{2}}{\hbar \left(
\omega _{\eta _{k0}}^{2}-\omega ^{2}-2i\gamma _{0}\omega \right)
}E_{0}e^{-i\left( \omega t+\sigma _{q_{\eta }}^{l,r}\right) +\gamma
_{0}t}
\end{equation}
and we can introduce the oscillator strengths of the normal modes of
vibrations into (\ref{24}) by the relation
\begin{equation}
\label{25}
f_{q_{\eta _{k0}}}=\frac{2m\omega _{\eta _{k0}}\left| \left< \eta _{k}\left|
q_{\eta }\right| \eta _{0}\right> \right| ^{2}}{\hbar }.
\end{equation}
Now the induced dipole moments can be expressed as
\begin{equation}
\label{26}
d_{q_{\eta }}^{l,r}=\sum_{k}\left( a_{q_{\eta }}^{l,r}\right)
^{2}\frac{e^{2}f_{q_{\eta _{k0}}}}{m}\frac{E_{0}e^{-i\left( \omega t+\sigma
_{q_{\eta }}^{l,r}\right) +\gamma _{0}t}}{\omega _{\eta
_{k0}}^{2}-\omega ^{2}-2i\gamma _{0}\omega }.
\end{equation}

We can get the mean polarizability $\chi _{q_{\eta }}^{l,r}$ of the
crystal volume unit taking the part of the
induced dipole moment (the part at electric field)
and multiplying it by effective number $N'$ of compound oscillators in a
volume unit.
If we suppose that the last single oscillator in any compound oscillator is the
first oscillator in the next compound oscillator on any helix then the
effective number $N'=N/2$ where $N$ is the number of single oscillators
in a volume unit \cite{RiV}. If we introduce $\chi _{q_{\eta }}^{l,r}$
into Drude - Sellmaier dispersion relation for refractive indices of the
crystals
\begin{equation}
\label{27}
\bar{n}_{l,r}^{2}-1=4\pi \sum_{\eta =1}^{3}\chi _{q_{\eta }}^{l,r}
\end{equation}
we get
\begin{equation}
\label{28}
\bar{n}_{l,r}^{2}-1=\frac{2\pi Ne^{2}}{m}\sum_{\eta
=1}^{3}\sum_{k}\left( a_{q_{\eta }}^{l,r}\right)
^{2}\frac{f_{q_{\eta _{k0}}}}{\omega _{\eta _{k0}}^{2}-\omega
^{2}-2i\gamma _{0}\omega }.
\end{equation}

If we use the eqs. (\ref{28}) and (\ref{12}) we can solve the
difference of quadrats of complex refractive indices for the left and
the right circularly polarized waves
\begin{eqnarray}
\label{29}
\lefteqn{ \bar{n}_{l}^{2}-\bar{n}_{r}^{2}=\frac{8\pi Ne^{2}}{m}\left( \alpha
^{2}+\beta ^{2}\right) \sin \theta \left( \phi _{l}+\phi
_{r}\right) }\nonumber \\
&&\times \sum_{k}\left[ -\frac{\frac{2\cos \theta
+A_{1}}{2+A_{1}^{2}}f_{q_{1_{k0}}}}{\omega _{1_{k0}}^{2}-\omega ^{2}-2i\gamma
_{0}\omega } +\frac{f_{q_{2_{k0}}}\cos \theta }{\omega _{2_{k0}}^{2}-\omega
^{2}-2i\gamma _{0}\omega }-\frac{-\frac{2\cos \theta
+A_{3}}{2+A_{3}^{2}}f_{q_{3_{k0}}}}{\omega _{3_{k0}}^{2}-\omega ^{2}-2i\gamma
_{0}\omega } \right]
\end{eqnarray}
and if we further use the relation (\ref{01}) for the complex rotatory
power in which we use $\phi _{l}+\phi _{r}=\omega
d(\bar{n}_{l}-\bar{n}_{r})/c$ and
$\bar{n}_{l}^{2}-\bar{n}_{r}^{2}=
(\bar{n}_{l}+\bar{n}_{r})(\bar{n}_{l}-\bar{n}_{r})$ we get
\begin{eqnarray}
\label{30}
\lefteqn{ \bar{\rho }(\omega )=\frac{4\pi Nde^{2}}{mc^{2}}\omega ^{2}\left( \alpha
^{2}+\beta ^{2}\right) \sin \theta
\cdot \sum_{k}\Biggl[ -\frac{\frac{2\cos \theta
+A_{1}}{2+A_{1}^{2}}f_{q_{1_{k0}}}}{\omega _{1_{k0}}^{2}-\omega ^{2}-2i\gamma
_{0}\omega }}\nonumber \\
&&+\frac{f_{q_{2_{k0}}}\cos \theta }{\omega _{2_{k0}}^{2}-\omega
^{2}-2i\gamma _{0}\omega }-\frac{\frac{2\cos \theta
+A_{3}}{2+A_{3}^{2}}f_{q_{3_{k0}}}}{\omega _{3_{k0}}^{2}-\omega ^{2}-2i\gamma
_{0}\omega } \Biggr].
\end{eqnarray}

Now we may find the real and imaginary part of $\bar{\rho }(\omega )$
that are from (\ref{01}) ORD and CD
\begin{eqnarray}
\label{31}
\lefteqn{ \mbox{Re}\bar{\rho }(\omega )=\rho (\omega )=A\omega ^{2}\cdot
\sum_{k}\Biggl[ -\frac{\frac{2\cos \theta
+A_{1}}{2+A_{1}^{2}}f_{q_{1_{k0}}}\left( \omega _{1_{k0}}^{2}-\omega
^{2}\right) }{\left( \omega _{1_{k0}}^{2}-\omega ^{2}\right)
^{2}+4\gamma _{0}^{2}\omega ^{2}}}\nonumber \\
&&+\frac{
f_{q_{2_{k0}}}\cos \theta \left( \omega _{2_{k0}}^{2}-\omega
^{2}\right) }{\left( \omega _{2_{k0}}^{2}-\omega ^{2}\right)
^{2}+4\gamma _{0}^{2}\omega ^{2}}-
\frac{\frac{2\cos \theta
+A_{3}}{2+A_{3}^{2}}f_{q_{3_{k0}}}\left( \omega _{3_{k0}}^{2}-\omega
^{2}\right) }{\left( \omega _{3_{k0}}^{2}-\omega ^{2}\right)
^{2}+4\gamma _{0}^{2}\omega ^{2}}\Biggr],
\end{eqnarray}
\begin{eqnarray}
\label{32}
\lefteqn{ \mbox{Im}\bar{\rho }(\omega )=\sigma (\omega )=2A\gamma
_{0}\omega ^{3}\cdot
\sum_{k}\Biggl[ -\frac{\frac{2\cos \theta
+A_{1}}{2+A_{1}^{2}}f_{q_{1_{k0}}}
}{\left( \omega _{1_{k0}}^{2}-\omega ^{2}\right)
^{2}+4\gamma _{0}^{2}\omega ^{2}}}\nonumber \\
&&+\frac{
f_{q_{2_{k0}}}\cos \theta }{\left( \omega _{2_{k0}}^{2}-\omega ^{2}\right)
^{2}+4\gamma _{0}^{2}\omega ^{2}}-
\frac{\frac{2\cos \theta
+A_{3}}{2+A_{3}^{2}}f_{q_{3_{k0}}}}{\left( \omega _{3_{k0}}^{2}-\omega ^{2}\right)
^{2}+4\gamma _{0}^{2}\omega ^{2}}\Biggr],
\end{eqnarray}
where $A=\frac{4\pi Nde^{2}}{mc^{2}}(\alpha ^{2}+\beta ^{2})\sin \theta$.

We know that in the case $|\omega _{0}^{2}-\omega ^{2}|^{2} >
4\gamma_{0}^{2} \omega^{2}$
(it means in
the frequency region far from dichroic frequencies) the ORD result
(\ref{31}) must pass into ORD result in the paper \cite{Vsv} where we
used the model of three undamped coupled oscillators. The ORD result in
the \cite{Vsv} is expressed in the form of three terms with defined
frequency dependencies. The first term has the frequency dependence of
the type $\omega ^{2}/(\omega _{0}^{2}-\omega ^{2})$ and it is known as the Drude
term. The second term contains dependence $\omega ^{2}/(\omega
_{0}^{2}-\omega ^{2})^{2}$ (the Chandrasekhar term) and the third one is
proportional to $\omega ^{2}/(\omega _{0}^{2}-\omega ^{2})^{3}$. These
terms frequency dependences are multiplied by the following coefficients
\begin{eqnarray}
\label{33}
K_{k0}^{(1)}&=&-\frac{2\cos \theta
+A_{1}}{2+A_{1}^{2}}f_{q_{1_{k0}}}+f_{q_{2_{k0}}}\cos \theta
-\frac{2\cos \theta
+A_{3}}{2+A_{3}^{2}}f_{q_{3_{k0}}},\nonumber \\
K_{k0}^{(2)}&=&\frac{\mu _{1}}{m}\biggl[ -\frac{2\cos \theta
+A_{1}}{2+A_{1}^{2}}A_{3}f_{q_{1_{k0}}}+\frac{\mu _{2}}{\mu
_{1}}f_{q_{2_{k0}}}\cos \theta \nonumber \\
&&-\frac{2\cos \theta
+A_{3}}{2+A_{3}^{2}}A_{1}f_{q_{3_{k0}}}\biggr], \\
K_{k0}^{(3)}&=&\left( \frac{\mu _{1}}{m}\right) ^{2}\biggl[ -\frac{\mu
_{2}}{\mu _{1}}\frac{2\cos \theta
+A_{1}}{2+A_{1}^{2}}A_{1}f_{q_{1_{k0}}}-2f_{q_{2_{k0}}}\cos \theta
\nonumber \\
&&-\frac{\mu _{2}}{\mu _{1}}\frac{2\cos \theta
+A_{3}}{2+A_{3}^{2}}A_{3}f_{q_{3_{k0}}}\biggr] .\nonumber
\end{eqnarray}
We can rewrite our results (\ref{31}) and (\ref{32}) in similar way and
we get more convenient results for the approximation of the ORD and CD
experimental data
\begin{eqnarray}
\label{34}
\rho (\omega )&=&A\omega ^{2}\cdot
\sum_{k}\left\{\frac{K_{k0}^{(1)}\left( \omega _{k0}^{2}-\omega
^{2}\right)}{\left (\omega _{k0}^{2}-\omega ^{2}\right) ^{2}+4\gamma
_{0}^{2}\omega ^{2}}+\frac{K_{k0}^{(2)}\left[ \left( \omega
_{k0}^{2}-\omega ^{2}\right) ^{2}-4\gamma _{0}^{2}\omega ^{2}\right]
}{\left[ \left( \omega _{k0}^{2}-\omega ^{2}\right) ^{2}+4\gamma
_{0}^{2}\omega ^{2}\right] ^{2}}\right.\nonumber \\
&&\left. +\frac{K_{k0}^{(3)}\left( \omega _{k0}^{2}-\omega ^{2}\right)
\left[ \left( \omega _{k0}^{2}-\omega ^{2}\right) ^{2}-12\gamma
_{0}^{2}\omega ^{2}\right] }{\left[ \left( \omega _{k0}^{2}-\omega
^{2}\right) ^{2}+4\gamma _{0}^{2}\omega ^{2} \right] ^{3}}\right\},
\end{eqnarray}
\begin{eqnarray}
\label{35}
\sigma (\omega )&=&2A\gamma _{0}\omega ^{3}\cdot \sum _{k}\left\{
\frac{K_{k0}^{(1)}}{\left( \omega _{k0}^{2}-\omega ^{2}\right)
^{2}+4\gamma _{0}^{2}\omega ^{2}}+\frac{2K_{k0}^{(2)}\left( \omega
_{k0}^{2}-\omega ^{2}\right) }{\left[ \left( \omega _{k0}^{2}-\omega
^{2}\right) ^{2}+4\gamma _{0}^{2}\omega ^{2} \right] ^{2}}\right.
\nonumber \\
&&+\left. \frac{K_{k0}^{(3)}\left[ 3\left( \omega _{k0}^{2}-\omega
^{2}\right) ^{2}-4\gamma _{0}^{2}\omega ^{2}\right] }{\left[ \left(
\omega _{k0}^{2}-\omega ^{2}\right) ^{2}+4\gamma _{0}^{2}\omega
^{2}\right] ^{3}}\right\}
\end{eqnarray}
where we have neglected the expressions that are multiplied by small
values $\mu
_{1}^{3}$, $\mu _{1}^{4}$, $\mu _{2}^{3}$, $\mu _{2}^{4}$ and also the
expressions that are multiplied by $\mu _{1}^{2}$ and $\mu _{2}^{2}$ in
the numerators of the first terms in compound brackets.
\section{Discussion}
\label{disc}
\hspace\parindent
We can see that our results (\ref{30}), (\ref{31}) and (\ref{32}) do not
contain the rotational strengths. But the nonzero values of rotational
strengths are the condition for the existence of the optically active
quantum transition and the results of OA are often expressed by using
these strengths (see ref. \cite{Deut}). We are going to show that our
results can also be expressed in this way.

The rotational strengths are defined as the imaginary part of the scalar
product of the electric and the magnetic transition dipole moments that
means
\begin{equation}
\label{36}
R_{k0}={\rm Im}\left( \langle 0\vert \hat{\vec d}\vert k\rangle
\cdot \langle k\vert \hat{\vec m}\vert 0 \rangle \right)
\end{equation}
where $|0\rangle $ is the ground state, $|k\rangle $ any excited state
and $\hat{\vec d}$, $\hat{\vec m}$ are the operators of the electric and
the magnetic dipole moments. We will compute the rotational strengths
in the normal coordinates in our case of three coupled oscillators
\begin{equation}
\label{37}
R_{q_{\eta _{k0}}}={\rm Im}\left( \langle \eta_{0}\vert
\hat{\vec{d}}_{q_{\eta }}\vert \eta_{k}\rangle \cdot \langle \eta_{k}\vert
\hat{\vec{m}}_{q_{\eta }}\vert \eta_{0}\rangle \right), \quad \eta =1,2,3.
\end{equation}

We have solved the normal components of the rotational strengths for the
model of three coupled oscillators in \cite{Vsv}. We have derived the
expressions
\begin{eqnarray}
\label{38}
R_{q_{1_{k0}}}&=&-\frac{\hbar e^{2}d \left( \alpha ^{2}+\beta ^{2}\right)
f_{q_{1_{k0}}}\sin \theta \left( 2\cos \theta +A_{1}\right) }{2mc\left(
2+A_{1}^{2}\right) },\nonumber \\
R_{q_{2_{k0}}}&=&\frac{\hbar e^{2}d \left( \alpha ^{2}+\beta ^{2}\right)
f_{q_{2_{k0}}}\sin \theta \cos \theta }{2mc},\\
R_{q_{3_{k0}}}&=&-\frac{\hbar e^{2}d \left( \alpha ^{2}+\beta ^{2}\right)
f_{q_{3_{k0}}}\sin \theta \left( 2\cos \theta +A_{3}\right) }{2mc\left(
2+A_{3}^{2}\right) }\nonumber
\end{eqnarray}
and we can see that the complex rotatory power (\ref{30}) can be
expressed as
\begin{equation}
\label{39}
\bar{\rho }(\omega )=\frac{8\pi N\omega ^{2}}{\hbar c}\sum_{\eta
=1}^{3}\sum_{k}\frac{R_{q_{\eta _{k0}}}}{\omega _{\eta _{k0}}^{2}-\omega
^{2} -2i\gamma _{0}\omega }.
\end{equation}

In the same way we can also express the ORD and CD results
\begin{equation}
\label{40}
\rho (\omega )=\frac{8\pi N\omega ^{2}}{\hbar c}\sum_{\eta
=1}^{3}\sum_{k}\frac{\left( \omega _{\eta _{k0}}^{2}-\omega ^{2}\right)
R_{q_{\eta _{k0}}}}{\left( \omega _{\eta _{k0}}^{2}-\omega ^{2}\right)
^{2}+4\gamma _{0}^{2}\omega ^{2}},
\end{equation}
\begin{equation}
\label{41}
\sigma (\omega )=\frac{16\pi N\omega ^{3}\gamma _{0}}{\hbar c}\sum_{\eta
=1}^{3}\sum_{k}\frac{
R_{q_{\eta _{k0}}}}{\left( \omega _{\eta _{k0}}^{2}-\omega ^{2}\right)
^{2}+4\gamma _{0}^{2}\omega ^{2}}
\end{equation}
and we can accept these formulas as the crystal analogs of the formulas
which are well known from the optical activity of molecules \cite{MoM} or
were recently derived for delocalized molecular aggregates \cite{Wag}.

The ORD and CD are often expressed in the dependence on wavelength in
practice. Because of this reason the ORD and CD formulas that are rewritten in
this dependence are also often used. We can
present several ORD and CD formulas that are suitable for theoretical
approximation of experimental data and that are based on the single
terms of the formulas (\ref{34}) and (\ref{35}). For ORD we get the formulas
\begin{eqnarray}
\label{42}
\rho (\lambda )&=&\sum_{k}\frac{K_{1k}^{(\rho )}
\left( \lambda^{2}-\lambda_{k0}^{2} \right)}
{\left( \lambda^{2}-\lambda_{k0}^{2} \right)^{2}+
\Gamma^{2}\lambda^{2}},\nonumber\\
\rho (\lambda )&=&\sum_{k}\frac{K_{2k}^{(\rho )}\lambda^{2}
\left[\left( \lambda^{2}-\lambda_{k0}^{2} \right)^{2}-
\Gamma^{2}\lambda^{2} \right]}
{\left[\left( \lambda^{2}-\lambda_{k0}^{2} \right)^{2}+
\Gamma^{2}\lambda^{2} \right]^{2}},
\\
\rho (\lambda )&=&\sum_{k}\frac{K_{3k}^{(\rho )}\lambda^{4}
\left( \lambda^{2}-\lambda_{k0}^{2} \right)
\left[\left( \lambda^{2}-\lambda_{k0}^{2} \right)^{2}-
3\Gamma^{2}\lambda^{2} \right]}
{\left[\left( \lambda^{2}-\lambda_{k0}^{2} \right)^{2}+
\Gamma^{2}\lambda^{2} \right]^{3}}
\nonumber
\end{eqnarray}
where $K_{1k}^{(\rho )}$, $K_{2k}^{(\rho )}$, $K_{3k}^{(\rho )}$ are the
constants defined only by the crystal parameters. The central dichroic
wavelengths
$\lambda _{k0}$ correspond to the transition frequencies $\omega _{k0}$ and
the constant $\Gamma $ contains the damping constant $\gamma _{0}$ of
single oscillators. It is evident that forms of all these constants
can be easy derived from (\ref{34}). The values of these constants are
computed by the approximation of experimental data by using the least squares
method.

We can see that the first formula in (\ref{42}) passes into well known
Drude formula for wavelengths very different from central dichroic
wavelength. The second formula passes into Chandrasekhar formula
\cite{Chand2} on the same conditions. It is also known that for example
the ORD of $\alpha $-quartz
is very well approximated by the sum of the Drude formula and the Chandrasekhar
formula with one central dichroic wavelength. The form of the first two
relations in (\ref{42}) was derived in \cite{Vys2} by means of the
classical model of two coupled oscillators but the form of constants
$K^{(\rho )}$ was different.

We are able to determine similar conclusions also for the relations by
means of which we can approximate the CD experimental data. The terms in
the relation (\ref{35}) can be rewritten into dependence on wavelengths
by using new constants as
\begin{eqnarray}
\label{43}
\sigma\left( \lambda \right)&=&\sum_{k}\frac{K_{1k}^{(\sigma )}\lambda}
{\left( \lambda^{2}-\lambda_{k0}^{2} \right)^{2}+
\Gamma^{2}\lambda^{2}}
\nonumber\\
\sigma\left( \lambda \right)&=&\sum_{k}\frac{K_{2k}^{(\sigma )}
\lambda^{3} \left( \lambda^{2}-\lambda_{k0}^{2} \right)}
{\left[\left( \lambda^{2}-\lambda_{k0}^{2} \right)^{2}+
\Gamma^{2}\lambda^{2}\right]^{2}}
\\
\sigma\left( \lambda \right)&=&\sum_{k}\frac{K_{3k}^{(\sigma )}
\lambda^{5}\left[3\left( \lambda^{2}-\lambda_{k0}^{2} \right)^{2}-
\Gamma^{2}\lambda^{2}\right]}
{\left[\left( \lambda^{2}-\lambda_{k0}^{2} \right)^{2}+
\Gamma^{2}\lambda^{2}\right]^{3}}.
\nonumber
\end{eqnarray}

The form of the first and the second relation was also derived in \cite{Vys2}
with exception of the form of all constant $K^{(\sigma )}$. We have made
the first calculations which should tell us the meaning of these
practical relations and above all the meaning of the third relations
in the eqs. (\ref{42}) and (\ref{43}). Now we can mention that the third
relation in (\ref{42}) has a very little meaning in the approximation of
the experimental ORD values of $\alpha$-quartz. If we approximate the
ORD values by means of all three relations (for one central dichroic
wavelength) we can essentially reduce the sum of least squares and
better approximate the ORD data mathematically but the value of
computed central dichroic wavelength is not probable. On the other hand
if we approximate the CD experimental data of sodium uranyl acetate
crystal (SUA) \cite{Dovh} by temperature $77 \hspace*{1mm} \mbox{K}$ we get much better
results by using all three relations (\ref{43}) with comparison to using
only the first two relations. The experimental data are very well described
by theoretical formula and the values of all constants are physically
real. But the practical meaning of relations (\ref{42}) and (\ref{43})
can be verified in the future by the approximation of ORD and CD
experimental data of more crystals.

\label{konec-vys1_99}

\end{document}